\documentclass[%
 aip,
 amsmath,amssymb,
 reprint,%
]{revtex4-1}

\usepackage{graphicx}
\usepackage{dcolumn}
\usepackage{bm}

\usepackage[utf8]{inputenc}
\usepackage[T1]{fontenc}
\usepackage{mathptmx}

\usepackage{listings}

\begin{document}

\title{Direct N-Body problem optimisation using the AVX-512 instruction set}

\author{J. Pedregosa-Gutierrez}
\email{jofre.pedregosa@univ-amu.fr}
\affiliation{Aix Marseille Univ, CNRS, PIIM, Marseille, France}

\author{J. Dempsey}
\affiliation{QuickThread Programming}%

\date{\today}

\begin{abstract}
The integration of the equations of motion of N interacting particles, represents a classical problem in many branches of physics and chemistry. The direct N-body problem is at the heart of simulations studying Coulomb Crystals. We present an hand-optimized code for the latest AVX-512 set of instructions that achieve a single core speed up of $\approx 340\%$ respect the version optimized by the compiler. The increase performance is due a optimization on the organization of the memory access on the inner loop on the Coulomb and, specially, on the usage of an intrinsic function to faster compute the $1/\sqrt{x}$. Our parallelization, which is implemented in OpenMP, achieves an excellent scalability with the number of cores. In total, we achieve $\approx 500GFLOPS$ using a just a standard WorkStation with one Intel Skylake CPU (10 cores). It represents $\approx 75\%$ of the theoretical maximum number of double precision FLOPS corresponding to Fused Multiplication Addition (FMA) operations.
\end{abstract}

\maketitle

\section{Introduction}
The N-body problem refers commonly to the situation where the motion of N point-like particles interact through a pair-wise force which depends only on their relative positions. Such problem is found in the context of astrophysical simulations~\cite{aarseth_nbody1_1999}, general molecular dynamics~\cite{mazur_common_1997} and one component plasmas~\cite{dubin_trapped_1999}, among others. Most of the published work regarding the optimization of the N-body problem is presented in the context of the astrophysical related topics~\cite{aarseth_nbody1_1999}, where a significant part of the research focus on which approximations can be made, which interactions can be neglected, etc. Another active field also exist where the N-body problem plays a crucial role: laser cooled ions in Radio Frequency (RF) traps. In such systems, the ions are confined in a Radio-Frequency (RF) trap or in a Penning trap~\cite{knoop_trapped_2016}, where the trapped particles are cooled down to mili-Kelvin temperatures. At such low temperatures, the ions form ordered structures, known as Coulomb Crystals~\cite{thompson_ion_2015}. In such scenario, the Coulomb interaction among all N particles have to be taken into account and the complexity of the problem is necessarily of O($N^2$). As the number of ions in such Coulomb Crystals is of the order of $10^1 - 10^4$, code optimized by the compiler has provided acceptable computing times for many situations~\cite{schiffer_phase_1993, zhang_molecular-dynamics_2007}.

However, a new range of interesting topics have emerged where computing time is greatly increased. For example, when studying the stopping power of a Coulomb Crystal~\cite{bussmann_stopping_2006}, where a large number of ions is needed, $N = 10^5$. Another situation where computing time can be an issue is in the study of the Kibble-Zurek Mechanism (KZM) using trapped ions, in which 1D to 2D topological transition is crossed non adiabatically, leading to the formation of structural defects~\cite{del_campo_structural_2010}. The KZM theory provides with a probability of defect generation, and therefore a large number of numerical experiments are needed in order to achieve statistically relevant results~\cite{pedregosa-gutierrez_defect_2020}. Reduced computation times could also help to wide-spread the use of synthetically generated CCD images to obtain the temperature of the crystal~\cite{rouse_superstatistical_2015}. Finally, a new type of charged particle detector has been proposed which uses the change on laser induced fluorescence on a Coulomb Crystal as a signal~\cite{poindron_non-destructive_2021}. Such type of studies require both, a large number of particles and a need for statistics.

The choice of the target architecture is important if a maximum of performance wants to be obtained. The introduction of CUDA in 2006~\cite{Nickolls_scalable_2008} by nVIDIA corporation was the birth of GPU accelerated computing which became popular quickly thanks to the large number of cores present in GPUs. However, in order to get close to the theoretical peak computing power, typically expressed in units of FLOPs (Floating Point Operations per second), a minimal number of particles is needed. For example, in a comparison test between different GPUs performed in 2013~\cite{capuzzo-dolcetta_performance_2013}, the nVIDIA Tesla K20, with an official Double Precision (DP) of 1170 GFLOPs, lead to $\approx 400$GFLOPs for $N=1024$ and $\approx 40$GFLOPs when $N=256$, well below the peak performance. Modern workstations have also improve over the years, and they can easily provide more than 1 DP TFLOPs for under 2k€. For example, an Intel® Xeon® W-2155 has a theoretical performance of 1263 DP GFLOPs.

Therefore, in order to benefit to a wider audience, we considered worth exploring the performance of a standard workstation for the case of N-Body problems. Of particular interest is the AVX-512 instructions set found in Intel CPUs as it allows of processing 8 DP numbers per clock and per core.

In the following, we present a discussion of how the standard, single thread, optimized N-body algorithm can be further improved using the AVX-512 instructions set. In particular, we will discuss: the role played by the different possible approximations to the numerical evaluation of $1/\sqrt{x}$ and their impact in terms of performance and the optimization of the locality of the data in order to reduce memory access, leading to virtually no cache misses at L1. It is followed by a presentation of our OpenMP implementation, which leads to an excellent scalability with the number of cores in the system tested.

\section{Molecular dynamics simulations of trapped ions}
The problem at hand consists then in integrate numerically the equations of motion of $N$ interacting ions in the presence of a trapping potential. In the linear RF trap case (with the trap axis along $z$), the equations of motion to solve are:

\begin{align}\label{eq:eq_of_motion}
\partial_{tt} x_i &= \frac{Q^2 k_C}{m}\sum_{j=1}^N{\frac{x_i - x_{j}}{|\vec{r}_{i}-\vec{r}_{j}+ \epsilon|^3}} - \frac{2 Q V_{RF} \cos{2 \pi \Omega t}}{m r_0^2} x_j - \frac{\Gamma}{m}\partial_t x_i \nonumber\\
\partial_{tt} y_i &= \frac{Q^2 k_C}{m}\sum_{j=1}^N{\frac{y_i - y_{j}}{|\vec{r}_{i}-\vec{r}_{j}+ \epsilon|^3}} + \frac{2 Q V_{RF} \cos{2 \pi \Omega t}}{m r_0^2} y_j - \frac{\Gamma}{m}\partial_t y_i \nonumber \\
\partial_{tt} z_i &= \frac{Q^2 k_C}{m}\sum_{j=1}^N{\frac{z_i - z_{j}}{|\vec{r}_{i}-\vec{r}_{j}+ \epsilon|^3}} - \omega^2 z_i - \frac{\Gamma}{m}\partial_t z_i 
\end{align}

where $\vec{r} = (x,y,z)$, $k_C = \frac{1}{4\pi\epsilon_0}$ is the Coulomb's constant, $Q$ and $m$ are the charge and the mass of the trapped ions, $\Omega$ and $V_{RF}$ are the frequency and voltage amplitude at which the trap is operated. In order to simulate laser cooling, a friction term, $\Gamma$, has been introduced. For a more detailed information about ion trap dynamics, we recommend~\cite{knoop_trapped_2016}.

Several differences can be noticed respect the astrophysics case: the trapped species are generally of the same type, i.e: same charge and mass, implying less operations needed in the algorithm. There exist situations were multiple species are simultaneously trapped for different purposes, like for quantum logic spectroscopy~\cite{schmidt_spectroscopy_2005} or for cold chemistry studies~\cite{willitsch_chemical_2008}. However, the number of different species is usually limited to two, meaning the problem can be separated in three N-body problems, the interaction among the species 1, interaction among specie 2 and interaction among 1 and 2. In the present study, one single specie has been used.

The choice of the numerical integration method is also important as it has implications on what it is called one "time-step". While gravitational codes tend to use the fourth-order Hermite integrator~\cite{makino_hermite_1992}, MD codes tend to use the Velocity-Verlet algorithm~\cite{zhang_molecular-dynamics_2007}, which is the one used in the present work. The Velocity-Verlet algorithm, with constant time step, can be reduced to three steps:
\begin{itemize}
\item Update position: $ r_{i+1} = r_{i} + v_{i} dt + \frac{dt^2}{2} a_{i} $
\item Compute acceleration at the new position: $ a_{i+1}(r_{i+1})$
\item Update velocity: $v_{i+1} = v_{i} + \frac{dt}{2} \left( a_i + a_{i+1} \right)$
\item Update acceleration for next iteration: $a_{i+1} \to a_i$
\end{itemize}

In our case, the acceleration, eq\ref{eq:eq_of_motion}, includes also a friction term that depends on the velocity and it is use to simulate laser cooling. Therefore the total energy of the system is not conserved. Nevertheless, it has been used successfully to reproduce experimental results, validating such methodology. 

\section{The N-body interaction }
A classical implementation for the computation of all the pair forces, written in C language, is given by:

\begin{lstlisting}[language=C]
for(size_t i = 0; i < n_ions; i++ ){
    a_x[i] = 0.0e00;
    a_y[i] = 0.0e00;
    a_z[i] = 0.0e00;
    for(size_t j = 0; j < n_ions; j++ ){
        rji[0]  = r_x[i] - r_x[j]; 
        rji[1]  = r_y[i] - r_y[j]; 
        rji[2]  = r_z[i] - r_z[j]; 

        r2inv   = rji[0]*rji[0]
                + rji[1]*rji[1]
                + rji[2]*rji[2]
                + softening;
                
        r2inv   = 1.0e0/sqrt(r2inv);
        r2inv   = r2inv * r2inv
                * r2inv * alpha;

        a_x[i] += rji[0]*r2inv;
        a_y[i] += rji[1]*r2inv;
        a_z[i] += rji[2]*r2inv;
    }
}
\end{lstlisting}

In the context of the AVX-512 instructions set, a 512-bit wide vector is processed simultaneously by each core. This implies that the optimal number of ions must be a multiple of 8, when using Double Precision.

Assuming that $N$ is a multiple of 8 and that the arrays are properly aligned in memory, we could expect that the Intel compiler optimizes the assembly code such that it takes advantage of the AVX-512 instructions when the -O3 -march=native flags are used on a CPU which implements such set of instructions. In particular the use of the Fused Multiplication Adition (FMA), which perform $z = xy + c$ in one assembly instruction.

By looking at the assembly code generated by the compiler, FMA instructions are correctly used, but it has decided to use 256-bit vector (4 DP numbers) instead of 512-bit vector. Nevertheless, it shows to which length modern compilers are able to optimize code.

A detailed analysis of the code performance highlights the bottleneck that represents the $1/\sqrt{x}$. The compiler documentation~\cite{Intel_IntrinscGuide} indicates that a 31 CPU clocks are required to compute the square root and 23 CPU clocks for the division for the type of CPU used in this work (Intel's SkyLake architecture). Both values refer to a mathematical operation on a 512-bit vector. For comparison, a multiplication, a sum and a FMA, need just 4 CPU clocks each. The AVX-512 instructions set includes a specific instruction to compute the $1/\sqrt{x}$ directly, instead of computing first the square root, and then the division. Such instruction needs only 9 CPU clocks. However, such reduction in CPU clocks gain (respect the 54 CPU clocks of the SQRT+DIV) comes with a loss of accuracy. Indeed, such instructions guarantees a maximum relative error of less than $2^{-14}$, while the SQRT+DIV guarantees a maximum relative error of less than $2^{-53}$. The consequence of such loss of accuracy is discussed next.

\section{Regarding the inverse square root}
The $1/\sqrt{x}$ term is the main single contributor to the total computational time of the N-body pair force computation. The inverse square root operation is common in many situations and well known techniques exist to compute fast approximations~\cite{libessart_scaling-less_2017}. Once an initial estimation of $y = 1/\sqrt{x}$ is available, the Newton-Raphson (NR) method can be used to improve the approximation. Each iteration of the NR method doubles the number of bits of precision~\cite{libessart_scaling-less_2017}. The estimation of $y_n$ of $1/\sqrt{x}$ is obtained after $n$ iterations of :

\begin{equation}
    y_{i+1} = \frac{y_i}{2} ( 3 - x y_i^2)
\end{equation}

The AVX-512 instruction set includes the RSQRT14 instruction, which provides an initial approximation with 14 bits of precision. One iteration of the NR method, leads to 28 bits and two iterations should lead to 56 bits, however, the double precision format limits to a maximum of 53 bits of precision.
The overhead for each N-W iteration can be reduced to 3 multiplications and one FMA. So (RSQRT14 + 1 N-W) implies 25 CPU clocks vs the 54 CPU clocks for (SQRT+DIV).

In order to asses the impact of the precision on the calculation of $1/\sqrt{x}$, we will use the simple case of two ions in a 1D harmonic oscillator of secular frequency $\omega$. In this scenario, the acceleration on the first ion is given by:
\begin{align}
    a_c(x_1) &= -\omega^2 x + \frac{Q^2 k_C}{m}(x_2 - x_1)\left( R_{21}\right)^3
\end{align}
where $R_{21} = \frac{1}{\sqrt{(x_2 - x_1)^2}}$.

Assuming the use of the Velocity-Verlet algorithm, the error in the new position due to an error only on the $R_{21}$ term is :
\begin{equation}
    \Delta x_{i+1} = \frac{3}{2} dt^2 \frac{Q^2 k_C}{m} x_{21}\left( R_{21}\right)^3 \epsilon
\end{equation}
where $\epsilon = \frac{\Delta R_{21}}{R_{21}} = 2^{-14}$.

Now, taking $R_{21} = 1/d$, where $d = |x_2 - x_1|$ leads to :
\begin{equation}
    \Delta x_{i+1} = \frac{3}{2} dt^2 \frac{Q^2 k_C}{m} d^{-2} \epsilon
\end{equation}

The two body collision theory tell us that the minimum distance between two particles, in 1D, is given by:
\begin{equation}
    d_{min} = \frac{4 Q^2 k_C}{m v^2} = \frac{4 Q^2 k_C}{k_B T}
\end{equation}

Therefore, the relative error at the smallest distance, $d=d_{min}$ is given by:
\begin{equation}
    \frac{\Delta x_{i+1}}{d}  = \frac{3}{8} dt^2 \frac{k_B T}{m} \epsilon
\end{equation}

Similarly, the error in a velocity update due only to the error on the $R_{21}$ term, is:
\begin{equation}
    \Delta v_{i+1} = 3 dt \frac{Q^2 k_C}{m} d^{-2} \epsilon
\end{equation}

where we have assume that $a_{i+1} \approx a_{i}$. Assuming $d_{min}$ and a maximal velocity given by $v_{max}= \sqrt{k_B T / m}$, leads to:
\begin{equation}
    \frac{\Delta v_{i+1}}{v} = \frac{3}{16}\frac{dt}{Q^2 k_C} \sqrt{\frac{k^3_B T^3}{m}} \epsilon
\end{equation}

Table~\ref{tab:table1} gives some values for $\frac{\Delta x_{i+1}}{d_{min}}$ when two Ca$^+$ ions and a time step of $dt = 5ns$ are used. The value of $T=0.5mK$ correspond to the Doppler limit of the Ca$^+$ ion~\cite{foot_atomic_2004}.
\begin{table}[h!]
  \begin{center}
    \caption{Your first table.}
    \label{tab:table1}
    \begin{tabular}{c|c|c|c|c }
    Method   & $\epsilon$ &   T   & $\frac{\Delta x_{i+1}}{d}$ & $\frac{\Delta v_{i+1}}{v}$\\
    \hline
    rinvsqrt & $2^{-14}$  & 300K  & $7\cdot 10^{-4 }$& $3\cdot 10^{-4 }$ \\
    rinvsqrt & $2^{-14}$  &   1K  & $3\cdot 10^{-11}$& $5\cdot 10^{-8 }$ \\
    rinvsqrt & $2^{-14}$  & 0.5mK & $3\cdot 10^{-21}$& $6\cdot 10^{-13}$ \\
    \hline
    rinvsqrt + 1NW & $2^{-28}$   & 300K  & $4\cdot 10^{-8 }$& $2\cdot 10^{-8 }$ \\
    rinvsqrt + 1NW & $2^{-28}$   &   1K  & $2\cdot 10^{-15}$& $3\cdot 10^{-12}$ \\
    rinvsqrt + 1NW & $2^{-28}$   & 0.5mK & $2\cdot 10^{-25}$& $4\cdot 10^{-17}$
    \end{tabular}
  \end{center}
\end{table}

The above results are a simplification of the N-body case, but they provide a valuable estimation of the expected errors.

\section{Coulomb Crystals structures comparison}
In order to compare the effect of the accuracy level on realistic scenarios, relevant for Coulomb Crystal related studies, we have introduced a friction term to the acceleration, $-\frac{\Gamma}{m}\vec{v}$, in order to simulated the laser cooling.

\begin{figure}
\includegraphics[width=1.0\columnwidth]{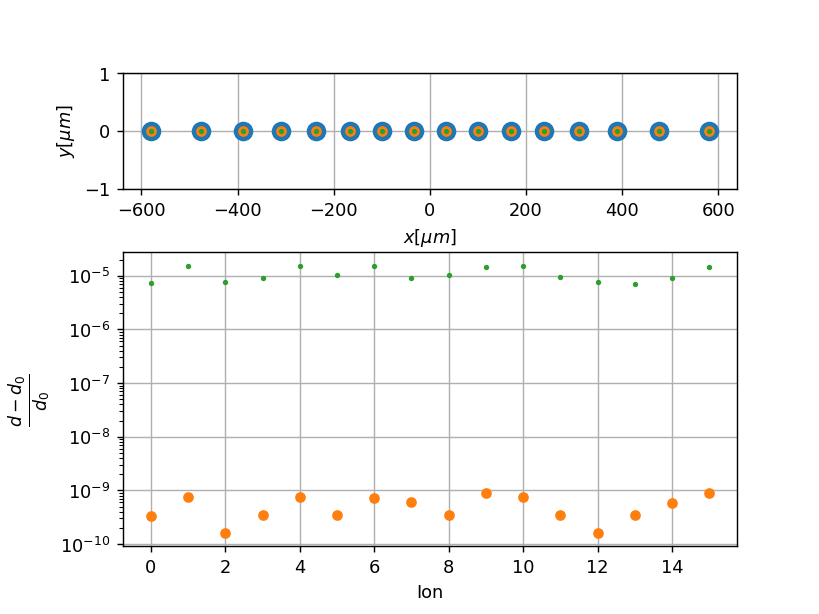}
\caption{\label{fig:chain} TOP: Comparison of final ion distribution of a simulated ion chain when using three different methods to compute the $1/\sqrt{x}$ term: "SQRT+DIV" (large blue dots), "RSQRT14" (medium orange dots) and "RSQRT14 + 1 NW" (small green dots). Bottom: the relative difference between "SQRT+DIV" and "RSQRT14+1NW" (large orange dots) and between "SQRT+DIV" and "RSQRT14" (small green dots) for each particle is shown.}
\end{figure}

A small system of $N=16$ Ca$^+$ ions in a configuration leading to an ion chain has been generated, see figure~\ref{fig:chain}. The other parameters of the simulation are: $\Omega / 2\pi = 2MHz$, $r_0=2.5mm$, $V_{RF} = 20V$, $\omega/2\pi = 1.581kHz$ and $\Gamma = 10^{-19}kg/s$. The same values of $\Omega$ and $r_0$ will be used through this paper. Three different ways to compute the $1/\sqrt{x}$ term have been tested: "SQRT+DIV", "RSQRT14" and "RSQRT14 + 1NW". After $5\cdot 10^3$ time-steps,

The ion distribution, see figure~\ref{fig:chain}~Top, is indistinguishable at this scale. The relative difference respect the  "SQRT+DIV", measured as $\bar{d} = \left|  \frac{d - d_{0}}{d_0}  \right|$, for each particle is plotted in figure~\ref{fig:chain}~Bottom, showing indeed that the "RSQRT14 + 1NW" leads to a closer result than the "RSQRT14".

The average of the relative difference leads to $\bar{d} =  10^{-5}$ for "RSQRT14" and $\bar{d} =  5\cdot10^{-10}$ for "RSQRT14+1NW". However, for many practical situations, the results obtained with "RSQRT14" are sufficient.

\begin{figure}
\includegraphics[width=1.0\columnwidth]{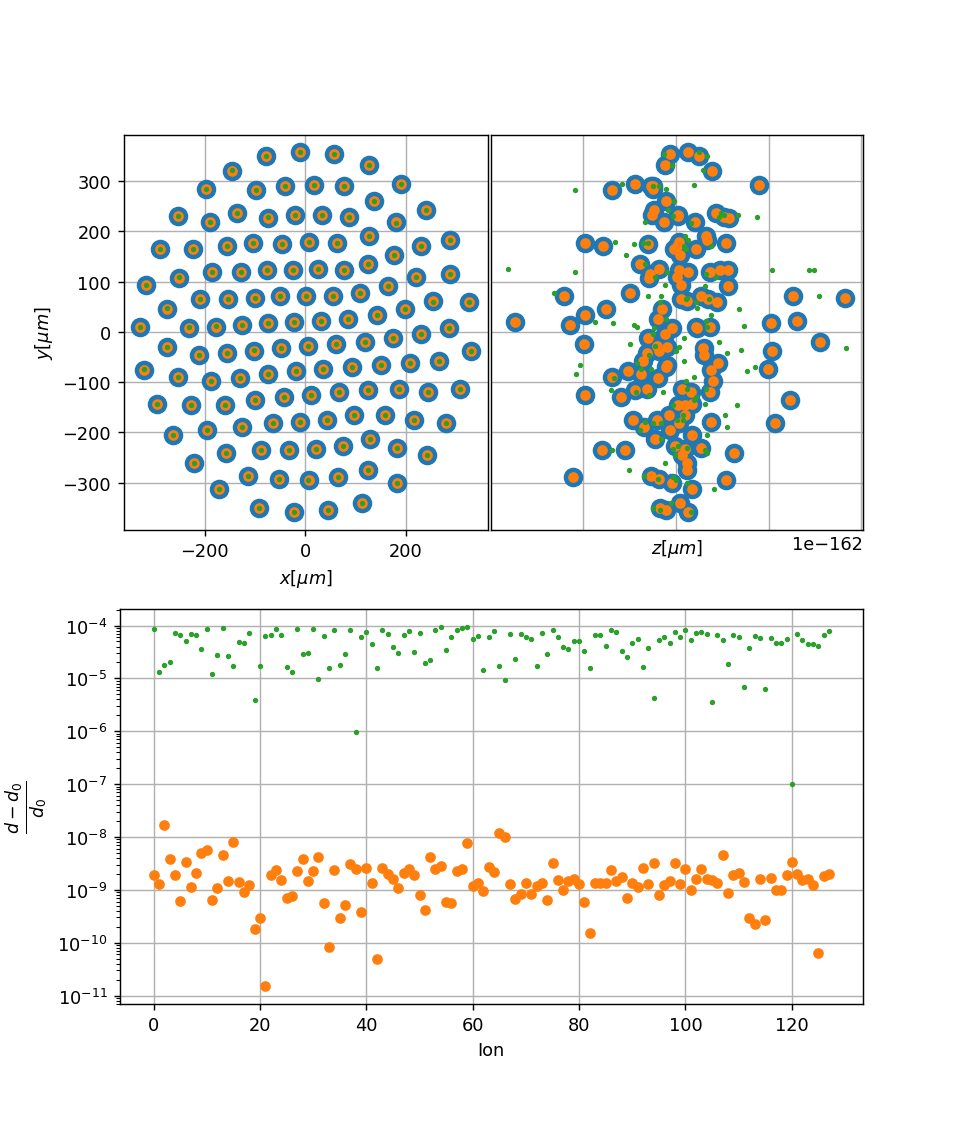}
\caption{\label{fig:pancake} Top: Comparison of final ion distribution for a 2D crystal when using "SQRT+DIV" (barge blue dots), "RSQRT14" (medium orange dots) and "RSQRT14 + 1 NW" (small green dots). Bottom: the relative difference between "SQRT+DIV" and "RSQRT14+1NW" (large orange dots) and between "SQRT+DIV" and "RSQRT14" (small green dots) for each particle is shown.}
\end{figure}

Another system consisting of a 2D crystal along the radial plane has been generated with $N=128$, see figure~\ref{fig:pancake}. The RF voltage and the axial secular frequency used were $V_{RF} = 352V$, $\omega/2\pi = 1.581kHz$ and $\Gamma = 10^{-21}kg/s$. The two top graphs of figure~\ref{fig:pancake} show the final spatial configuration after $10^{6}$ time-steps which agrees with the expected results~\cite{dubin_trapped_1999}. The agreement between the "SQRT+DIV" and "RSQRT14+1NW" is excellent and the agreement with "RSQRT14" can be considered acceptable in many situations. The bottom graph of figure~\ref{fig:pancake} indicate the relative error respect the "SQRT+DIV" for each ion's position.

\section{Performance}
In the precedent sections we reported on the different methods to obtain the inverse square root, and analysed, with a simple model and with some examples, the effect of the different precisions. In the following, the achieved performance in terms of FLOPs and the time taken for one time-step of the Velocity-Verlet algorithm is reported. Only intrinsic functions using packed 512-bit vector instructions have been used through the integration kernel.

\subsection{Minimisation of data access}
An important aspect of the new code is the minimisation of the data access. In the "classical" double loop, the interaction of the particle $i$ (outer loop) with the all the $N$ ions is done sequentially. By using the 512-Bytes packed structure (8 double precision), it is possible to compute the interaction of the $i$ particle on the outer loop with 8 particles in the inner loop simultaneously.

Things can be further improved by computing the 8 interactions on the $i$ loop and re-using the 8 particles in the $j$ loop by introducing an extra inner loop. At the end of this new inner loop the particles are re-arrange. In this way the interaction of the contribution of the 8 particles from the inner loop to the force of the 8 particles on the $i$ loop have been computed. In pseudocode, it can be written as :
\begin{lstlisting}[language=C]
for(size_t i = 0; i < n_ions; i+=8 ){
    for(size_t j = 0; j < n_ions; j+=8 ){
        for (size_t k = 0; k < 8; k++) {
            compute_ri_packed_rj_packed
            rotate_rj_packed
        }}}
\end{lstlisting}

This approach saves N/8 memory access which compensates largely the cost of the re-arrangement of the $r_j$ packed structure. The performance results presented in the following make use of such optimized approach.

We have compared five possible ways of computing $1/\sqrt(x)$. We have included the "INVSQRT" function which has the same accuracy as "SQRT+DIV". The reported performance metrics that follow, have been obtained using a Intel Xeon W-2155 with HyperThreading deactivated. It has a single core theoretical maximum performance of 62 GFLOPS for DP addition, so 124 GFLOPs for DP FMA, assuming that 512-bit vector are used.

\subsection{Single thread performance}
\begin{figure}
\includegraphics[width=1.0\columnwidth]{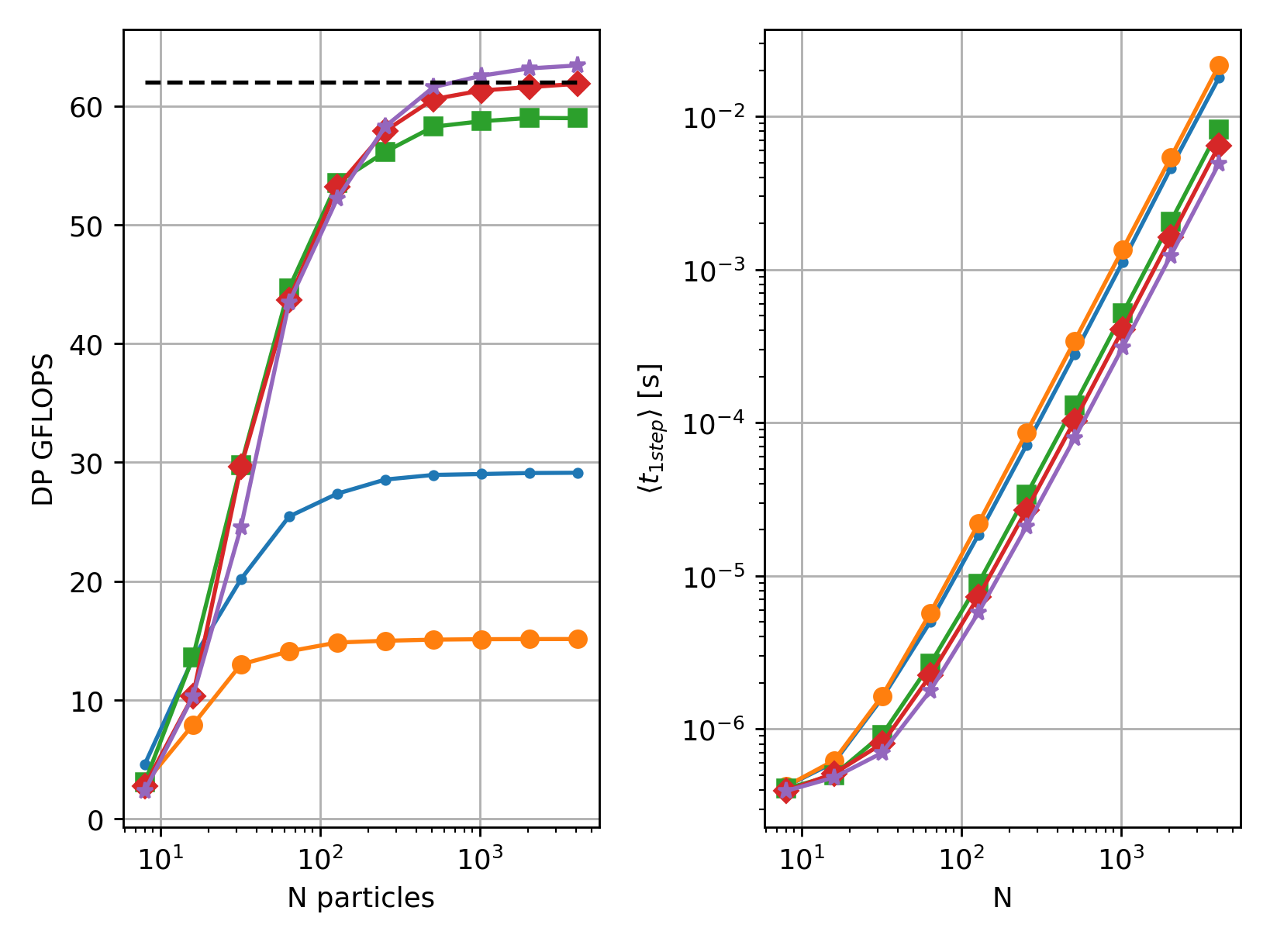}
\caption{Result for a single core using different methods to compute the $1/\sqrt(x)$ term. Left: Number of DP GFlops obtained vs the number of particles. Right: average time needed to perform one time step on the integration algorithms vs the number of particles used. Large orange dots: "SQRT+DIV", small blue dots: "INVSQRT", green squares: "RSQRT14+2NW", red diamonds: "RSQRT14+1NW", purple stars: "RSQRT14". }
\label{fig:Single_thread_performance} 
\end{figure}

While the best single thread performance does not necessarily means the best performance once the code has been parallelized over the available cores, it is nevertheless a good starting point. Figure~\ref{fig:Single_thread_performance}-left shows the evolution of the DP GFLOPs with the number of simulated ions. The horizontal dashed line indicated the 62GFLOPs DP addition threshold for the particular CPU used. The number of FLOPs obtained for small numbers is significantly lower, clearly indicating that a minimum number of particles is needed to obtain good performance. Figure~\ref{fig:Single_thread_performance}-right shows the time needed to perform a full simulation, consisting on $10^6$ time-steps, divided by the number of integration steps. Indeed, a large value of FLOPS only indicates the number of floating operations per second, however a badly implemented code can provide a large value of FLOPS and yet take a long time to perform one step of the integration algorithm due to unnecessary/redundant calculations.

\begin{figure}
\includegraphics[width=1.0\columnwidth]{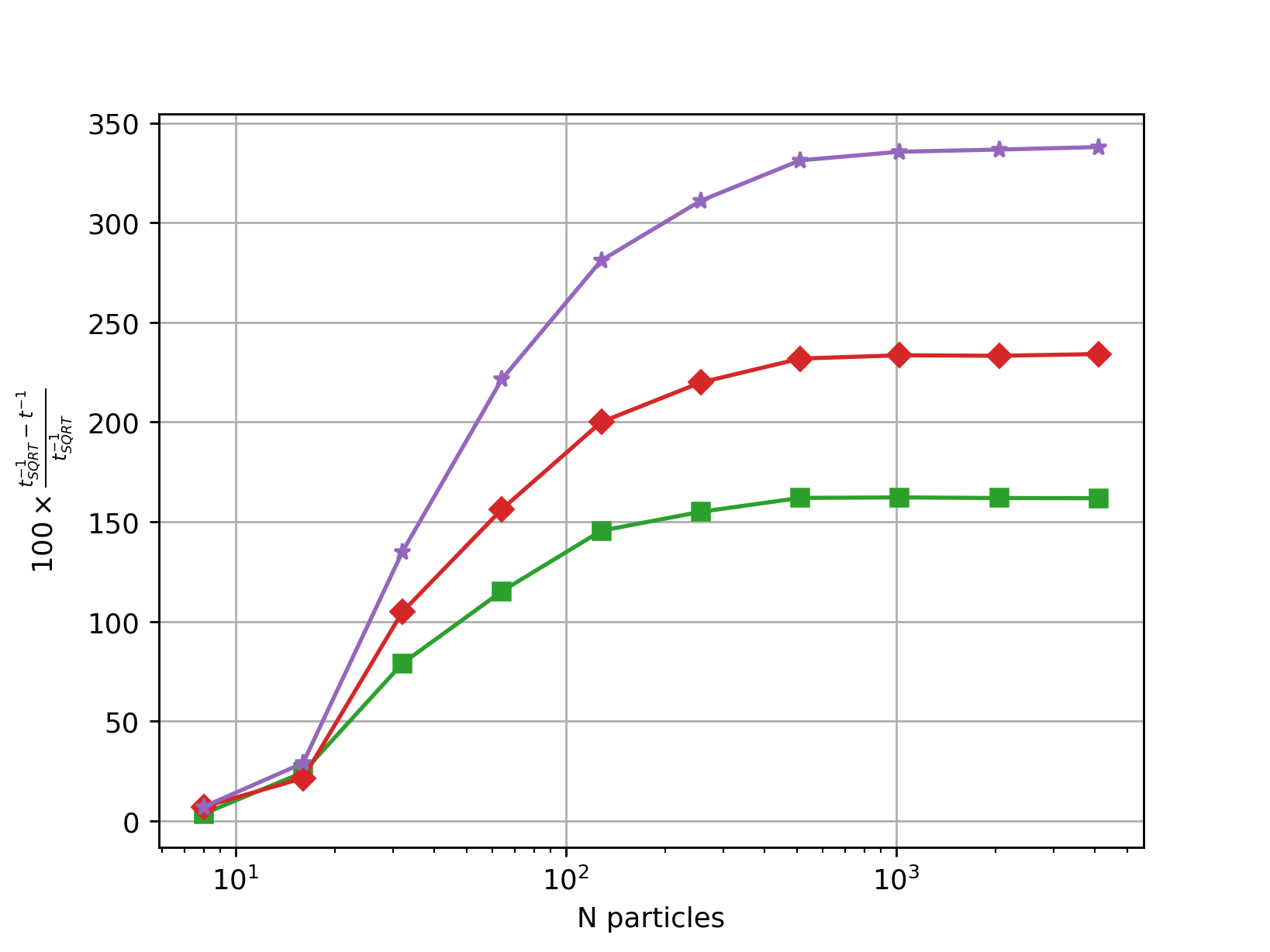}
\caption{Single core relative acceleration respect the "SQRT+DIV" method vs the number of particles used. Green squares: "RSQRT14+2NW", red diamonds: "RSQRT14+1NW", purple stars: "RSQRT14".}
\label{fig:Single_thread_Accel_rs_INVSQRT}
\end{figure}

The time scales as $N^2$ very quickly, indicating that it is indeed the Coulomb interaction which dominates the execution time. In order to fully grasp the increase on performance, figure~\ref{fig:Single_thread_Accel_rs_INVSQRT} shows the acceleration respect "SQRT+DIV" defined as $100\times\frac{t^{-1}_{SQRT}  - t^{-1}}{t^{-1}_{SQRT}}$. Notice the use of average time of one time step rather than the GFLOPS values, which would have lead to a misleading acceleration definition. We insist, from a practical point of view, it is the time that it takes to get a result that matters, not the FLOPS. Our results show a speed up of $\approx 166\%$ respect the standard double precision accuracy and an acceleration gain of $\approx 340\%$ when using the lowest accuracy on the $1/\sqrt(x)$ term. Our understanding of the reason for the increase in performance for similar accuracy is because of the out-of-order capabilities achieved thanks to the different Scheduler Ports of modern processors~\cite{intel_skylake}. When "SQRT" and "DIV" are called, they introduce a dependency chain, and therefore the compiler has to compute them sequentially. Moreover, there is only one port in the Skylake microarchitecture~\cite{intel_skylake} that can perform "DIV" and "SQRT", which is probably introducing a bottleneck. By splitting the calculation explicitly on the source code, the compiler can dispatch the different operations among the available ports to be done in a more optimize way.

\subsection{Multi core performance}
The code has been parallelized using OpenMP. The approach consist in manually imposing which ions are handle by each thread. The total number of ions must be a multiple of 8 and multiple of the number of threads used. If a number of particles does not satisfy such condition, dummy particles need to be introduced in order to keep performance. In such way, the simulation is fully reproducible as OpenMP has no freedom to dispatch the particles and therefore the order on which the calculations are performed should be the same from run to run. In our implementation, only one parallel section is used, which contains the integration loop. In such a way, each core perform the integration of $N_{ions}/n_{thread}$ particles.

\begin{figure}
\includegraphics[width=1.0\columnwidth]
{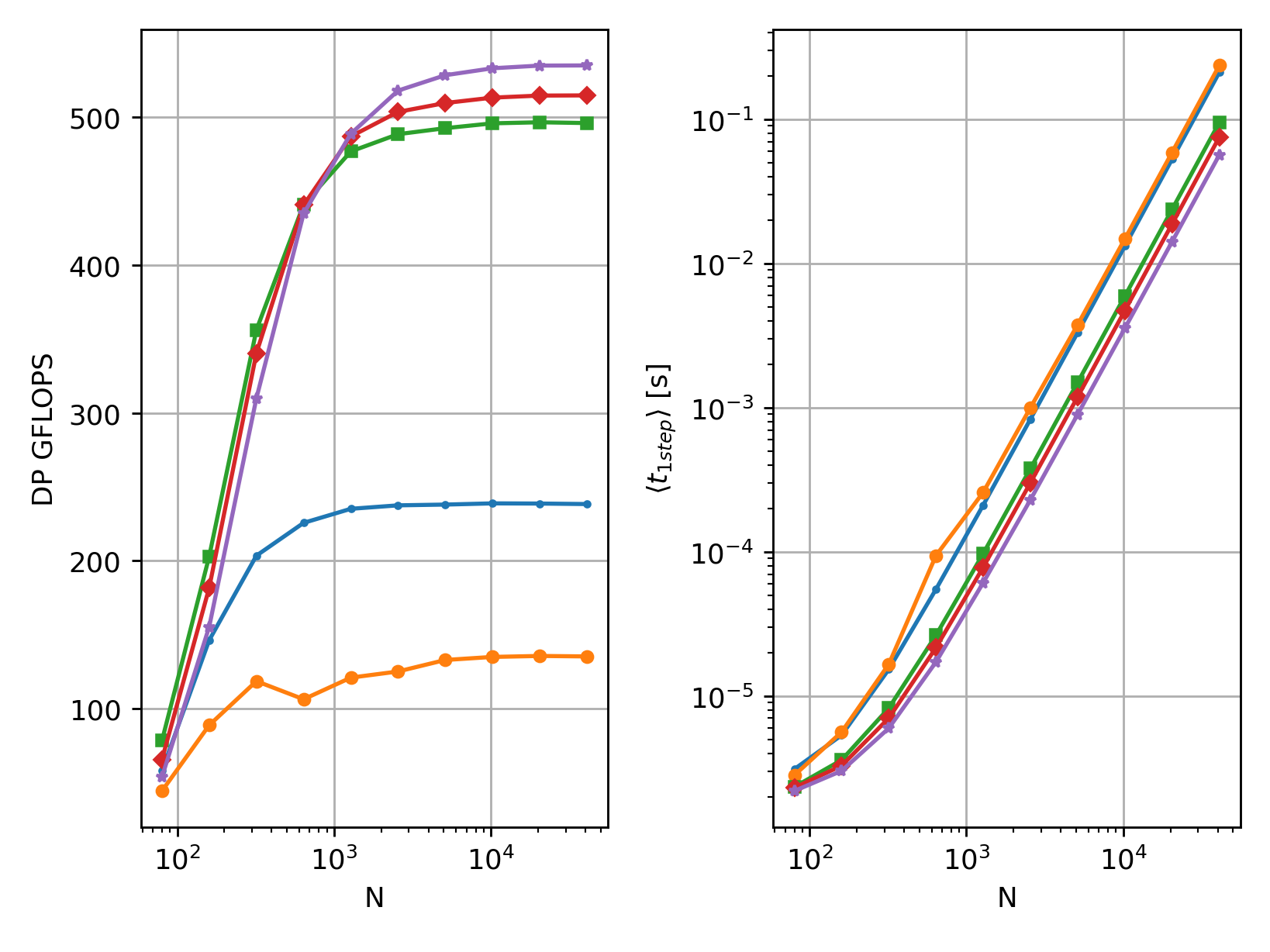}
\caption{Left: Value of the DP GFLOPS using the 10 cores of the CPU achieved in function of the total number of simulated particles. Right: Averaged time needed for 1 time step in function of the number of simulated particles. Large orange dots: "SQRT+DIV", small blue dots: "INVSQRT", green squares: "RSQRT14+2NW", red diamonds: "RSQRT14+1NW", purple stars: "RSQRT14".}
\label{fig:OpenMP}
\end{figure}

\begin{figure}
\includegraphics[width=1.0\columnwidth]
{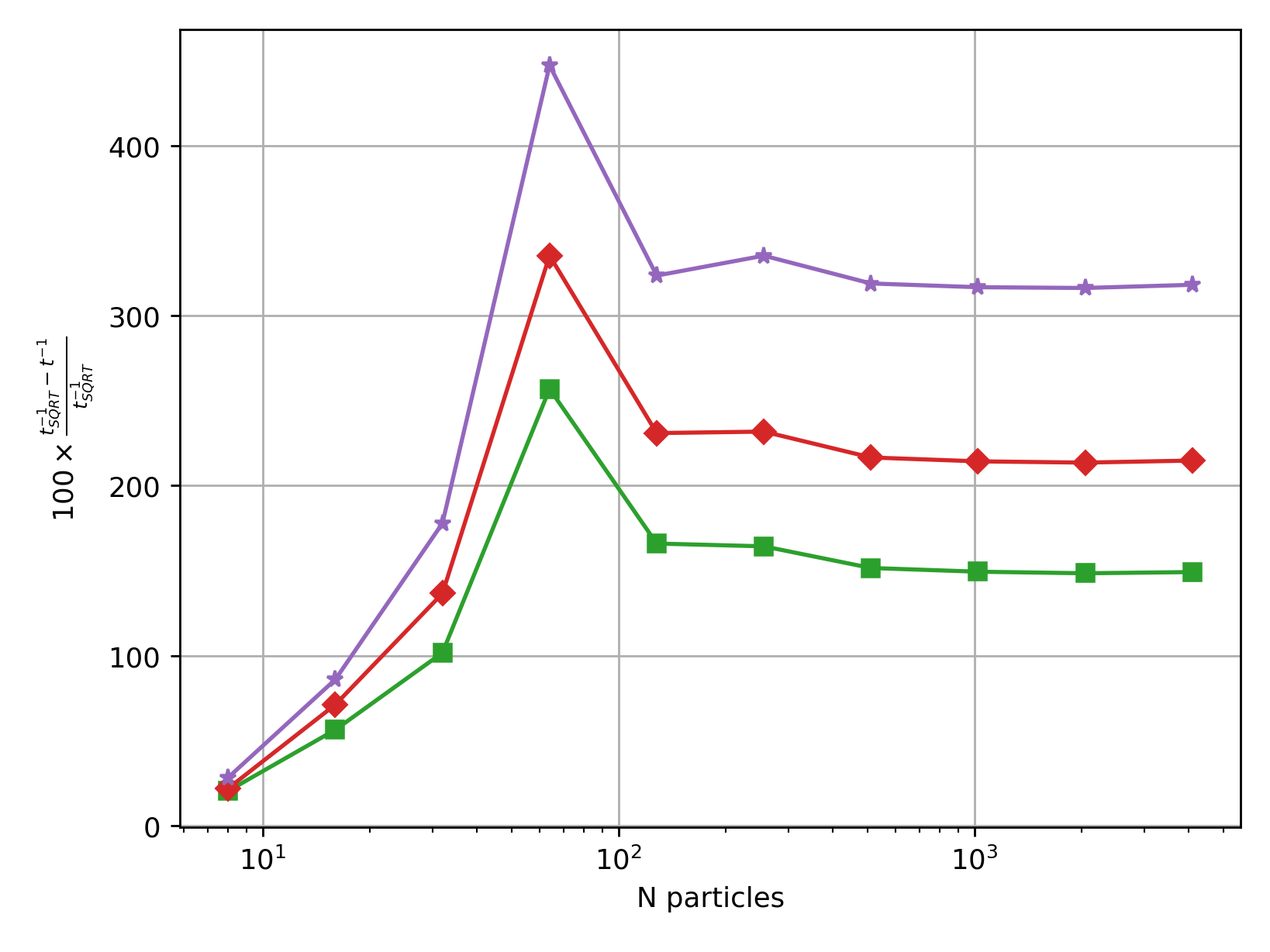}
\caption{Relative acceleration respect the "SQRT+DIV" method vs the number of particles using the OpenMP version of the code. Green squares: "RSQRT14+2NW", red diamonds: "RSQRT14+1NW", purple stars: "RSQRT14".}
\label{fig:OpenMP_Increase}
\end{figure}

The results of the performance of our implementation are shown in figure\ref{fig:OpenMP}. The results have been obtained using the 10-cores available at our Intel Xeon W-2155 workstation (running Ubuntu 16.04.7). The relative increase in performance respect "SQRT+DIV", is slightly lower than the single core case as shown in figure~\ref{fig:OpenMP_Increase}. Nevertheless, an increase of $\approx 150\%$ at equal accuracy and of $\approx 320\%$ for the lowest accuracy. , which is the option taken by the compiler, is $\approx 300 \%$.

\section{Conclusion}
In summary, we have shown how the algorithm to compute the N-body interaction can be speed up roughly by $320\%$ if a loss on accuracy on the calculation of the inverse of the square root is acceptable. If the standard double precision is required, our implementation provides still a significant $140\%$ increase on performance. Such gain in performance is the combination of using a particular method to compute the $1/\sqrt{x}$, which uses a low accuracy approximation of the $1/\sqrt{x}$ coupled with 2 Newton-Raphson iteration to recover the standard level of accuracy, and the explicit use of low level operation of 512byte packed arrays. While the time invested to improve the algorithm was significant, the results reported in the present work illustrates how, if a particular type of code is used extensively, hand tune the implementation of the actual algorithm by going to low level programming can bring significant performance gains.


\end{document}